\documentclass[runningheads]{llncs}

\usepackage[T1]{fontenc}
\usepackage{graphicx}
\usepackage{amsmath}
\usepackage{marvosym} 
\usepackage[
    colorlinks=true,
    linkcolor=blue,
    citecolor=blue,
    urlcolor=black,
    bookmarks=true,
    bookmarksnumbered=true
]{hyperref}
\hypersetup{bookmarksdepth=subsection}
\begin{document}
\title{
    Real-Time Interactive Hybrid Ocean: Spectrum-Consistent 
    Wave Particle-FFT Coupling
    }
\titlerunning{SCWP-FFT}
%
\author{Shengze Xue\textsuperscript{(\Letter)} \and
Yu Ren \and
Jiacheng Hong \and
Run Ni \and
Shuangjiu Xiao\textsuperscript{(\Letter)}\and 
Deli Dong\textsuperscript{(\Letter)}
}
\authorrunning{Xue et al.}
%
\institute{Digital ART Lab, Shanghai Jiao Tong University, Shanghai, China
\email{\{richardxue,xiaosj,arli@sjtu.edu.cn\}}\\
\url{http://dalab.se.sjtu.edu.cn/www/home/}
}
\maketitle              
\begin{abstract}
Fast Fourier Transform-based (FFT) spectral oceans are 
widely adopted for their efficiency and large-scale 
realism, but they assume global stationarity and 
spatial homogeneity, making it difficult to represent 
non-uniform seas and near-field interactions (e.g., ships and floaters). 
In contrast, wave particles capture local wakes and ripples, 
yet are costly to maintain at scale and hard to match global spectral statistics.

We present a real-time interactive hybrid ocean: a 
global FFT background coupled with local wave-particle (WP) 
patch regions around interactive objects, jointly driven 
under a unified set of spectral parameters and dispersion. 
At patch boundaries, particles are injected according to 
the same directional spectrum as the FFT, aligning the local 
frequency-direction distribution with the background and 
matching energy density, without disturbing the far field.

Our approach introduces two main innovations: (1) Hybrid ocean representation. 
We couple a global FFT background with local WP patches under a 
unified spectrum, achieving large-scale spectral consistency 
while supporting localized wakes and ripples.(2) Frequency-bucketed 
implementation. We design a particle sampling and GPU-parallel 
synthesis scheme based on frequency buckets, which preserves 
spectral energy consistency and sustains real-time interactive performance.
Together, these innovations enable a unified framework that 
delivers both large-scale spectral realism and fine-grained 
interactivity in real time.

\keywords{computer animation  \and ocean simulation \and sea wave spectrum
\and real-time simulation \and wave particle \and fast fourier transform.}
\end{abstract}

\section{Introduction}
In virtual reality, visual effects, maritime training, and 
digital twin applications, realistic, dynamic, and interactive 
ocean-surface simulation is essential. Real oceans exhibit 
strong spatiotemporal variability and non-uniformity—shaped 
by wind seas, swell, depth, and other factors\cite{stewart2008introduction}—and include localized 
disturbances such as ship wakes and floater ripples. These 
macro-micro features jointly challenge visual immersion and 
physical plausibility.

State-of-the-art large-scale ocean synthesis commonly adopts 
spectral, FFT-based methods\cite{flugge2017realtime,mastin2007fourier}, 
which superpose directional-spectrum waves in the frequency 
domain to generate periodic surfaces, achieving efficient, 
globally consistent results for film and games. However, 
like Gerstner-wave stacking\cite{fournier1986simple}, 
FFT methods assume spatial homogeneity and global stationarity, 
offering limited local control and struggling to reproduce ship 
wakes, floater ripples, shallow-water effects, and other localized 
phenomena with precision. By contrast, the Wave Particle approach 
introduced by Yuksel\cite{yuksel2007wave} and refined by Jeschke 
et al.\cite{jeschke2018water,jeschke2017water} targets real-time 
water interaction rather than large-area synthesis: it propagates 
wave particles with physical attributes to flexibly generate 
locally non-uniform disturbances from moving objects, excelling 
at high-fidelity near-field detail. Yet maintaining vast particle 
counts at scale is costly, and the resulting local field is 
difficult to keep consistent with the global ocean in energy 
distribution and spectral statistics. More complex particle-only 
methods such as PBF\cite{macklin2013position} and 
SPH\cite{muller2003particle} are even more expensive and typically 
limited to small-scale offline rendering.

This situation yields an ``impossible triangle'' in 
large-scale ocean simulation: statistical models like 
FFT scale efficiently\cite{tessendorf2001simulating}; 
wave-based particle methods enable local interaction 
and non-uniformity; but simultaneously achieving real-time 
performance, global consistency, and fine near-field 
interaction remains unresolved.

To address this, we propose a real-time interactive hybrid 
ocean. An FFT ocean serves as the global background wave 
field, while wave particle patch regions are introduced 
around interactive objects. The two representations share 
unified spectral parameters and dispersion; particles are 
injected at patch boundaries according to the same spectrum 
as the background, ensuring local spectral and energy 
distributions match the global field. With frequency 
bucketing and parallel reconstruction, the wave particle 
energy field is efficiently composed into height maps and 
smoothly fused with the global FFT ocean, yielding seamless 
transitions between near and far fields. Our method resolves
 this ``impossible triangle'' by combining the efficiency of 
 FFT with the interactivity of WP under a consistent 
 spectrum. Specifically, our contributions are:
\paragraph{Hybrid ocean representation.}
A novel WP–FFT coupling under a unified spectrum that integrates 
large-scale stability with localized interactions.  
\paragraph{Frequency-bucketed implementation.}
A new wave particle sampling and GPU-parallel synthesis scheme that 
maintains spectral consistency while enabling real-time performance.

\section{Previous Work}
Wave simulation methods and ocean modeling methods both reproduce 
water-surface dynamics but differ in focus. Wave simulation centers 
on generating wave fields via spectral, particle, or fluid formulations 
with prescribed statistics or dynamics, largely independent of scene 
geometry and boundary conditions, emphasizing plausibility and efficiency. 
Ocean modeling emphasizes integrated behavior under specific boundaries, 
coupling global and local representations across resolutions so that 
one domain shows far-field state and near-field interaction.

\subsection{Wave Simulation Methods}
FFT-based spectral methods superpose directional-spectrum components 
in the frequency domain and invert to height fields, yielding efficient, 
large-scale, globally consistent oceans\cite{flugge2017realtime,mastin2007fourier}. 
They leverage spectra such as 
Phillips\cite{newell2008role} and JONSWAP\cite{mazzaretto2022global}, 
are GPU-friendly, and power systems like Crest\cite{bowles2017crest}, 
but assume spatial homogeneity and global stationarity, limiting local 
control and fidelity for wakes, ripples, and shallow-water effects. 
wave particle methods\cite{jeschke2018water,jeschke2017water,yuksel2007wave} 
target real-time interaction, generating locally non-uniform disturbances 
from moving objects, yet become costly at scale and struggle to stay 
consistent with global spectral statistics. Pan et al.\cite{pan2025simulation} 
combine spectral control with wave particles for local control, but 
remain fundamentally particle-based and thus insufficient for real-time, 
very large domains.

\subsection{Ocean Modeling Methods}
Huang\cite{huang2021ships} couples high-resolution 
FLIP\cite{agrotis2016fluid} in the near field with a far-field 
BEM\cite{kythe2020introduction}, exchanging boundary data 
bidirectionally. The framework delivers physically consistent 
transitions and convincing visuals but inherits the high cost of 
3D particle fluids, limiting real-time scalability to large oceans.
\bigskip
\\In summary, spectral methods are efficient but lack local control; 
wave particle methods enable interaction but do not scale; and full 
3D fluid couplings are too expensive for real time. We therefore 
propose a real-time interactive hybrid ocean that fuses a global 
FFT background with local wave particle regions under unified 
spectral parameters, preserving large-scale frequency characteristics 
while achieving detailed, physically continuous interactions.

\section{Method}
This paper presents a real-time interactive hybrid ocean that 
couples a globally spectral, FFT-driven ocean with local 
Lagrangian wave particle patch regions. In this model, 
the primary wave field is synthesized via Fast Fourier 
Transform to ensure spectral consistency and computational 
efficiency at large scales, while detailed disturbance zones 
near interactive objects (e.g., ships, floaters) are reconstructed 
with wave particles to realize local non-uniformity and 
interactivity. To guarantee a seamless transition and physical 
consistency between the two representations, all wave particles 
are initialized from the same spectral parameters as the FFT 
and are generated under a statistical energy-density constraint, 
so that their energy expression and visual appearance remain 
compatible with the background wave field.

\subsection{Hybrid Ocean Representation}
The proposed hybrid ocean comprises two complementary wave 
representations: (i) a global background wave field 
synthesized by a spectral FFT, producing a large-scale, continuous, 
spectrum-consistent surface; and (ii) local wave particle patch 
regions placed around interactive objects to reconstruct non-uniform, 
disturbance-responsive near-field waves. The two are coupled at 
the parameter level by sharing a unified directional 
spectrum and dispersion relation.

The FFT ocean takes a directional spectrum as input and composes 
periodic waves in the frequency domain, well suited to statistically 
steady far-field behavior. Because computation is concentrated in 
the frequency domain and the entire height map is updated by a 
single inverse transform, local control is difficult. In contrast, 
wave particles provide a Lagrangian description of propagation: 
each particle carries direction, frequency, and amplitude, and 
contributes locally via a kernel. This makes the particle model 
adaptable to localized disturbances, particularly in highly 
interactive zones such as wakes and crest interference.

To fuse the two consistently in space, we adopt a local replacement 
strategy: around each interactive object, a rectangular wave particle 
patch region is defined, the FFT representation is disabled inside, 
and the surface is provided by the wave particle field. At patch 
boundaries, particles are injected according to the same directional 
spectrum as the FFT background, yielding a smooth transition in 
energy distribution, directional structure, and visual appearance. 
To avoid double counting or depletion of energy, the patch contribution 
is composited only locally; the global height field is obtained by 
stitching the background with all patch regions.

In this project we adopt the JONSWAP spectrum as the global spectral 
model and use a multiplicative directional spreading:
\begin{equation}
  S_J(\omega,\theta) = S_J(\omega)\,Dir(\omega,\theta)
\end{equation}
where $\omega$ is the angular frequency, $\theta$ is the angle between 
wave and wind directions, $S_J(\omega)$ gives the frequency-domain energy 
distribution, and $Dir(\omega,\theta)$ is the directional spreading function.

Compared with the Phillips spectrum, JONSWAP is defined over the joint 
frequency-direction domain\cite{donatini2024physically}, better fitting 
measured ocean energy distributions and, via the peak-enhancement factor 
$\gamma$, capturing wind-sea growth over finite fetch and 
duration\cite{horvath2015empirical}. Its one-dimensional form is:
\begin{equation}
  S_J(\omega) = \alpha g^2 \omega^{-5}
  \exp\left[ -\frac{5}{4} \left( \frac{\omega_p}{\omega} \right)^4 \right] \gamma^r
\end{equation}
with
\begin{equation}
  r = \exp\left[ -\frac{(\omega - \omega_p)^2}{2\sigma^2 \omega_p^2} \right]
\end{equation}
where $\omega_p$ is the peak frequency, $\sigma$ controls the peak width, and 
$\gamma$ is the peak-enhancement factor. We use the standard parameterizations:
\begin{equation}
  \alpha = 0.076 \left( \frac{U_{10}^2}{F g} \right)^{0.22}, \quad 
  \omega_p = 22 \left( \frac{g^2}{U_{10} F} \right)^{1/3}, \quad 
  \gamma = 7.0 \left( \frac{gF}{U_{10}^2} \right)^{-0.142}
\end{equation}
where $U_{10}$ is the wind speed at 10 m height, $F$ is the fetch, 
and $g$ is gravitational acceleration.

We employ the commonly used \emph{cos-2s} spreading, and the function should be
normalized\cite{forristall1998worldwide}:
\begin{equation}
\mathrm{Dir}(\omega,\theta)=\frac{\Gamma\!\big(s(\omega)+1\big)}
{2\sqrt{\pi}\,\Gamma\!\big(s(\omega)+\tfrac12\big)}\!
\left[\cos\!\left(\tfrac{\theta}{2}\right)\right]^{2s(\omega)},
\quad \int_{-\pi}^{\pi}\mathrm{Dir}(\omega,\theta)\,\mathrm{d}\theta=1
\end{equation}

In the above, $\Gamma(\cdot)$ is the Gamma function used for normalization, 
and $s$ is the spreading parameter that controls directional concentration 
around the mean wave direction. Following the observations of 
Mitsuyasu et al.\cite{mitsuyasu1975observations}, we parameterize $s(\omega)$ as
\begin{equation}
s(\omega) = 16.0 \left(\frac{\omega}{\omega_p}\right)^{\mu},
\qquad
\mu =
\begin{cases}
5.0,  & \omega \le \omega_p,\\[2pt]
-2.5, & \omega > \omega_p,
\end{cases}
\end{equation}

This hybrid representation directly enables the coexistence 
of large-scale consistency and fine-grained local detail, 
as highlighted in our contributions.

With the spectral update of the FFT background field being a 
well-established procedure, the remainder of this method 
focuses on how to generate and drive wave particle patch 
regions under the same spectrum.

\subsection{Spectrum-Based Initialization of Single wave particle}
\begin{figure}[!htb]
  \centering
  \begin{minipage}[t]{0.48\linewidth}
    \centering
    \includegraphics[width=\linewidth]{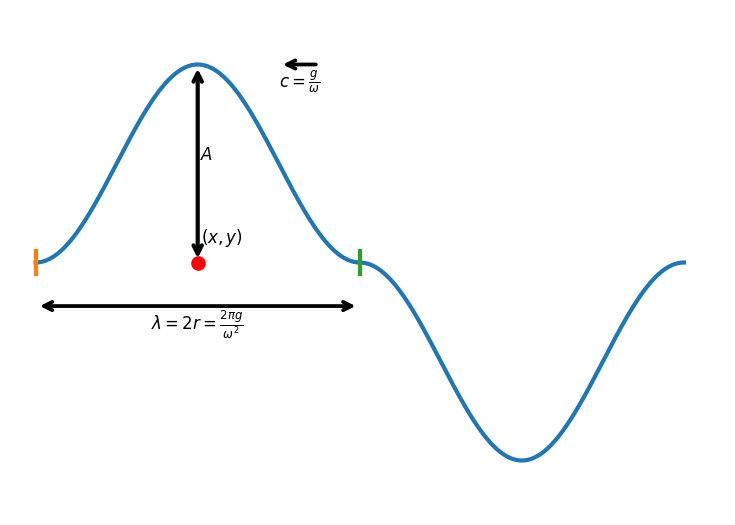}
    \medskip
    \small (a) Height distribution function of a single wave particle
  \end{minipage}\hfill
  \begin{minipage}[t]{0.48\linewidth}
    \centering
    \includegraphics[width=\linewidth]{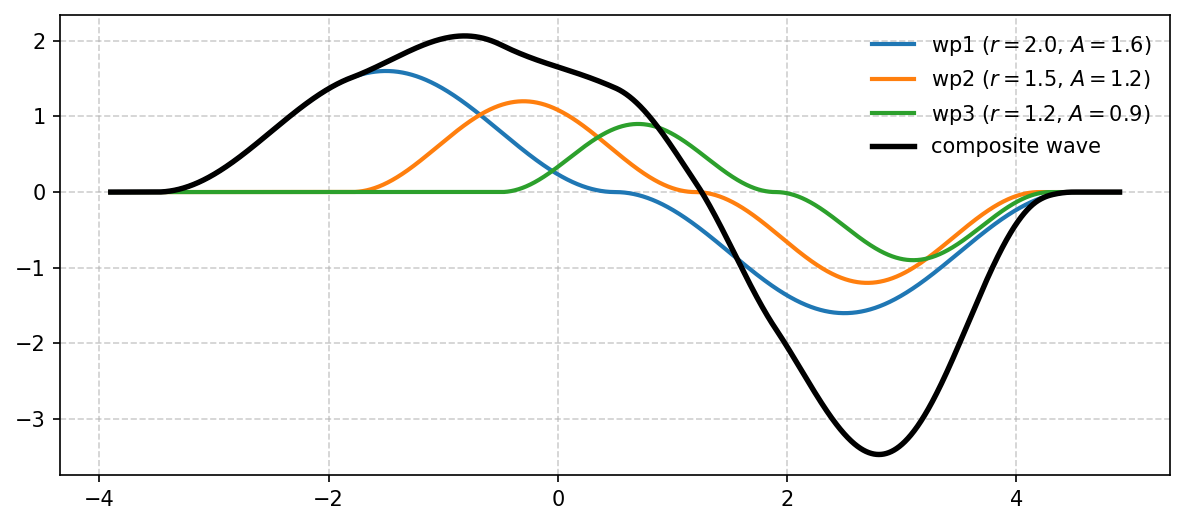}
    \medskip
    \small (b) Composited wave height distribution function
  \end{minipage}
  \caption{Wave height functions of a single wave particle 
  (including the positive-negative pair) and of a wave composed 
  of multiple particles.}
\label{fig:wave_pair}
\end{figure}

In the hybrid ocean model, the wave particle patch region must 
inherit consistent statistical characteristics from the spectrum 
to ensure continuity with the global FFT background in both energy 
structure and visual appearance. An intuitive illustration of how 
wave particles superpose into waves is shown in~\ref{fig:wave_pair}.
To this end, we adopt the spectrum-guided initialization proposed by 
Pan et al.\cite{pan2025simulation}, sampling each wave particle's 
attributes from a unified directional spectrum, including angular 
frequency $\omega$, propagation direction $\theta$, wavelength 
$\lambda$, amplitude $A$, and dispersion parameter $k$. This 
sampling process is aligned with the directional spectrum used 
by the FFT background, enabling spectrum-driven coherent evolution.

Particles are sampled over $(\omega,\theta)$. 
In deep water ($\omega^2=gk$, $k=2\pi/\lambda$)
\cite{newell2008role}, for fixed $\omega$:

\[
k=\frac{\omega^{2}}{g},\quad
\lambda=\frac{2\pi}{k}=\frac{2\pi g}{\omega^{2}}
\;\Rightarrow\;
r=\frac{\lambda}{2}=\frac{\pi g}{\omega^{2}},\quad
c=\frac{\omega}{k}=\frac{g}{\omega}.
\]
The particle amplitude $A$ can be computed from the spectral 
energy density function $S(\omega,\theta)$\cite{donatini2024physically}:

\begin{equation}
  A=\sqrt{2S(\omega,\theta)\Delta\omega\Delta\theta}
  \label{eq:A}
\end{equation}
Thus once $\omega$ and $\theta$ are given, $A$, $r$, and $c$ are uniquely determined, 
ensuring spectral energy consistency.

Within the hybrid ocean simulation framework, however, maintaining 
energy continuity between the local wave particle patch 
region and the global FFT background requires continuous 
particle injection at the patch boundaries. The key challenge 
lies in estimating an appropriate distribution of particle 
counts across frequencies $\omega$ and directions $\theta$ 
under given wind and directional spectrum conditions—particularly 
balancing the abundance of high-frequency short waves against the 
scarcity of low-frequency long waves.

\subsection{Spectrum-Based Estimation of wave particle Distribution}
The JONSWAP spectrum function $S(\omega,\theta)$ describes 
the distribution of ocean surface wave energy in the 
frequency-direction domain under given wind speed and wind 
direction. By integrating this spectrum over the entire 
frequency-direction plane, the total energy density per unit 
surface area can be obtained:
\begin{equation}
  \bar E = \int_{\omega_{\mathrm{min}}}^{\omega_{\mathrm{max}}}\int_{-\pi}^{\pi}
  S(\omega,\theta)\,\mathrm{d}\theta\ \mathrm{d}\omega
\end{equation}
\begin{figure}[!htb]
  \centering
  \begin{minipage}[t]{0.31\linewidth}
    \centering
    \includegraphics[width=\linewidth]{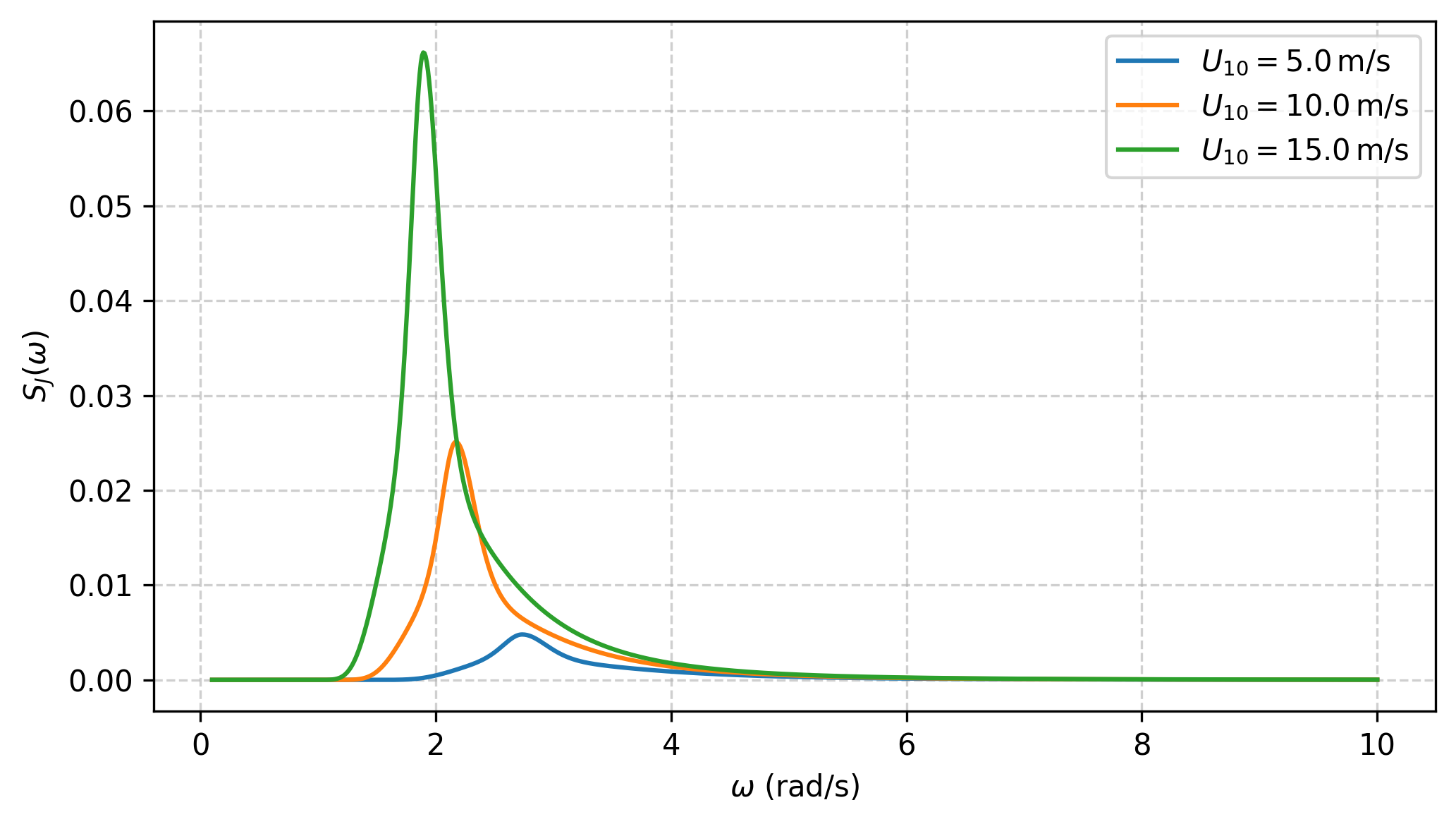}
    \medskip
    \small (a) JONSWAP spectrum under different wind speeds
  \end{minipage}\hfill
    \begin{minipage}[t]{0.35\linewidth}
    \centering
    \includegraphics[width=\linewidth]{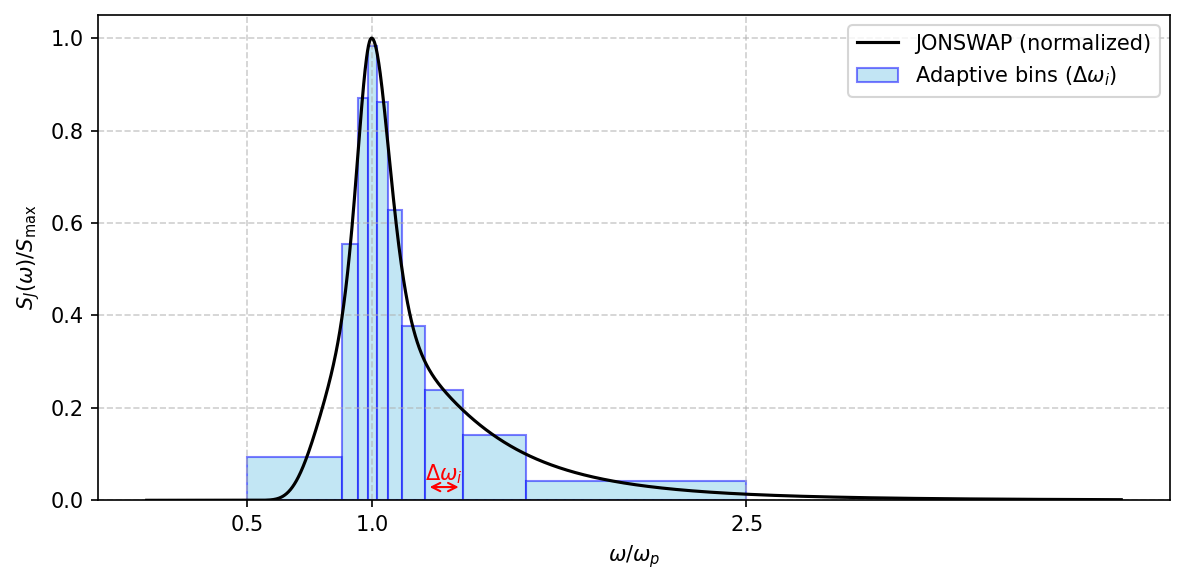}
    \medskip
    \small (b) The mathematical interpretation of adaptive $\bar E_{\omega_i}$ when
    $N_\omega=10$
  \end{minipage}
      \begin{minipage}[t]{0.3\linewidth}
    \centering
    \includegraphics[width=\linewidth]{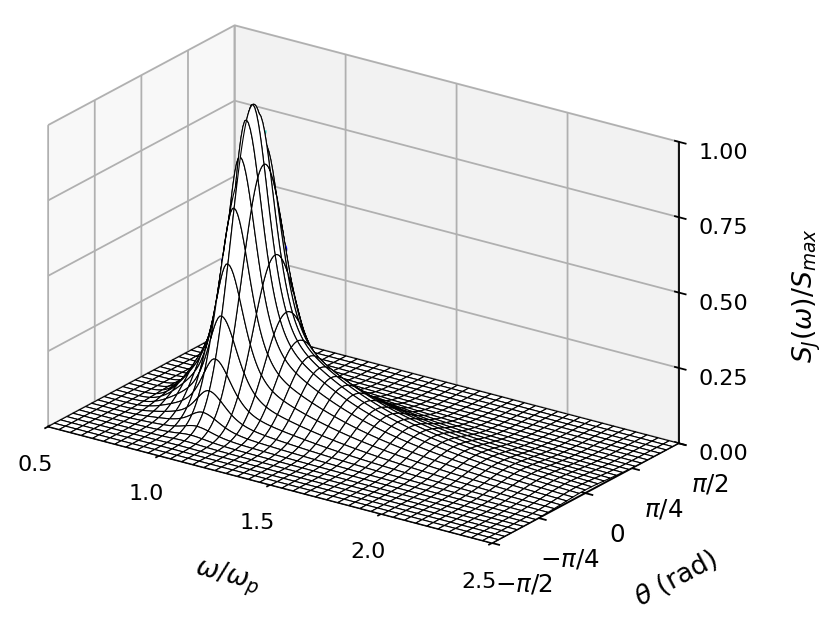}
    \medskip
    \small (c) 3D plot of the JONSWAP spectrum
  \end{minipage}
  \caption{Representation of the JONSWAP Spectrum and Sampling Strategy}
  \label{fig:dis_energy}
\end{figure}
In implementation, we define the sampling frequency range as 
$[0.5\omega_p\,,\,2.5\omega_p]$ to ensure efficiency and avoid 
noise\cite{Tian_Perlin_Choi_2011}. 
Instead of using a uniform step size,
we divide it into $N_\omega$ frequency buckets by equal-energy partitioning.
The sampling strategy is shown in Figure~\ref{fig:dis_energy}.
The total energy density of a frequency bucket containing all 
waves with frequency $\omega_i$ can then be expressed as:

\begin{equation}
  \bar E_{\omega_i} = \int_{-\pi}^{\pi}
  S(\omega_i,\theta)\,\mathrm{d}\theta\ \cdot \Delta\omega_i
\end{equation}
Here, $\Delta\omega_i$ is the bucket width, i.e., the sampling interval. 
Since all waves in the bucket share the same frequency ${\omega_i}$, 
their phase speed is uniquely determined as $c_i=\frac{g}{\omega_i}$. 
Over a boundary of length $L$, during a time step $\Delta t$, the 
area swept by waves of frequency $\omega_i$ is:

\begin{equation}
  \Delta S_{\omega_i} = L \cdot \Delta t \cdot c_i = L \cdot \Delta t \cdot \frac{g}{\omega_i}
\end{equation}
For simplicity, let the patch region be a square. When considering 
only net inflow energy, the sum over the four sides cannot be written as 
$4\Delta S$, since some waves propagate outward. However, inflow and 
outflow on opposite sides cancel out. Thus, for a square of side 
length $L$, the total swept area of incoming waves can be expressed as:

\begin{equation}
  \Delta S_{\omega_i,square} = 2 \cdot \Delta S_{\omega_i} = 2L \cdot \Delta t \cdot \frac{g}{\omega_i}
\end{equation}
Thus, over a time step $\Delta t$, the total incoming energy of waves with 
frequency ${\omega_i}$ into the region is:
\begin{equation}
\begin{split}
  \mathcal E_{\omega_i,square} 
  &= \rho g \cdot \Delta S_{\omega_i,square} \cdot \bar E_{\omega_i} \\
  &= 2\rho g L \cdot \Delta t \cdot \frac{g}{\omega_i} \cdot
  \int_{-\pi}^{\pi} S(\omega_i,\theta)\,\mathrm{d}\theta\ \cdot \Delta\omega_i
\end{split}
\label{eq:energy_1}
\end{equation}

On the other hand, from the wave particle perspective, 
the energy of a single deep-water wave particle with frequency $\omega_i$ 
and direction $\theta_j$ is:

\begin{equation}
\begin{split}
  \mathcal E_{\omega_i,\theta_j}
  &= \tfrac{1}{8}\,\rho g\,A_{\omega_i,\theta_j}^2 \,\pi r_i^2 \\
  &= \tfrac{1}{8}\,\rho g \cdot 2S(\omega_i,\theta_j)\Delta\omega_i\Delta\theta
     \cdot \pi\!\left(\frac{\pi g}{\omega_i^2}\right)^2 \\
  &= \tfrac{1}{4}\,\rho g\,\Delta\omega_i\,
     \frac{\pi^3 g^2}{\omega_i^4}\; S(\omega_i,\theta_j)\,\Delta\theta
\end{split}
\end{equation}

Wave particles sharing the same frequency $\omega_i$ but spanning all 
directions $\theta_j$ are referred to as a group.
The total energy of 1 group at frequency $\omega_i$ is:

\begin{equation}
 \mathcal E_{\omega_i,group} = \tfrac{1}{4}\,\rho g\,\Delta\omega_i\,
     \frac{\pi^3 g^2}{\omega_i^4}\; \sum_{j=1}^{N_{\theta}} S(\omega_i,\theta_j)\,\Delta\theta
\end{equation}

If within a time step $\Delta t$, the number of particle groups 
of frequency $\omega_i$ entering the square region is $N_i$, then:

\begin{equation}
  \begin{split}
  \mathcal E_{\omega_i,square} 
    &= N_i \cdot \mathcal E_{\omega_i,group} \\
    &= N_i \cdot \frac{1}{4}\,\rho g\,\Delta\omega_i\,
     \frac{\pi^3 g^2}{\omega_i^4}\; \sum_{j=1}^{N_{\theta}} S(\omega_i,\theta_j)\,\Delta\theta
  \end{split}
  \label{eq:energy_2}
\end{equation}
When $N_\theta$ is sufficiently large, the summation in 
(\ref{eq:energy_2}) can be approximated to:

\begin{equation}
  \sum_{j=1}^{N_{\theta}} S(\omega_i,\theta_j)\,\Delta\theta \approx
  \int_{-\pi}^{\pi} S(\omega_i,\theta)\,\mathrm{d}\theta
\end{equation}
By comparing equations(\ref{eq:energy_1}) and(\ref{eq:energy_2}), the number of particle groups with 
frequency $\omega_i$ entering the square region within a time 
step $\Delta t$ is:

\begin{equation}
  N_i = \frac{8L \cdot \Delta t \cdot \omega_i^3 }{\pi^3 g}
  \label{eq:result}
\end{equation}

Specifically, if the region is a rectangle with side 
lengths $L_1$ and $L_2$, then the number of particle groups  $N_i$ 
generated on opposite sides should correspond to two values:
\begin{equation}
  N_{i_1} = \frac{4L_1 \cdot \Delta t \cdot \omega_i^3 }{\pi^3 g},\quad
  N_{i_2} = \frac{4L_2 \cdot \Delta t \cdot \omega_i^3 }{\pi^3 g}
  \label{eq:result_2}
\end{equation}
Thus, we obtain the number of particles to be generated in 
each frequency bucket per time step under the energy consistency 
constraint.

It is worth noting that Eq.\~(\ref{eq:result},\ref{eq:result_2}) have clear 
physical meaning: its units reduce to a dimensionless particle 
group count, matching the derivation goal. The result is 
intuitive—particle groups scale with $\Delta t$ and $L$, 
and grow rapidly with frequency $\omega_i$, reflecting the 
abundance of short waves versus the rarity of long waves in 
real seas. Although $N_{\omega}$ and $N_{\theta}$ are decoupled 
from particle counts, this is reasonable: increasing $N_{\theta}$ 
raises particle numbers but lowers individual amplitudes, 
preserving energy. Thus the injection rule is both mathematically 
consistent and physically sound.

\section{Implementation}
\subsection{Pipeline Overview}
\begin{figure}[!htb]
  \centering
  \includegraphics[width=\linewidth]{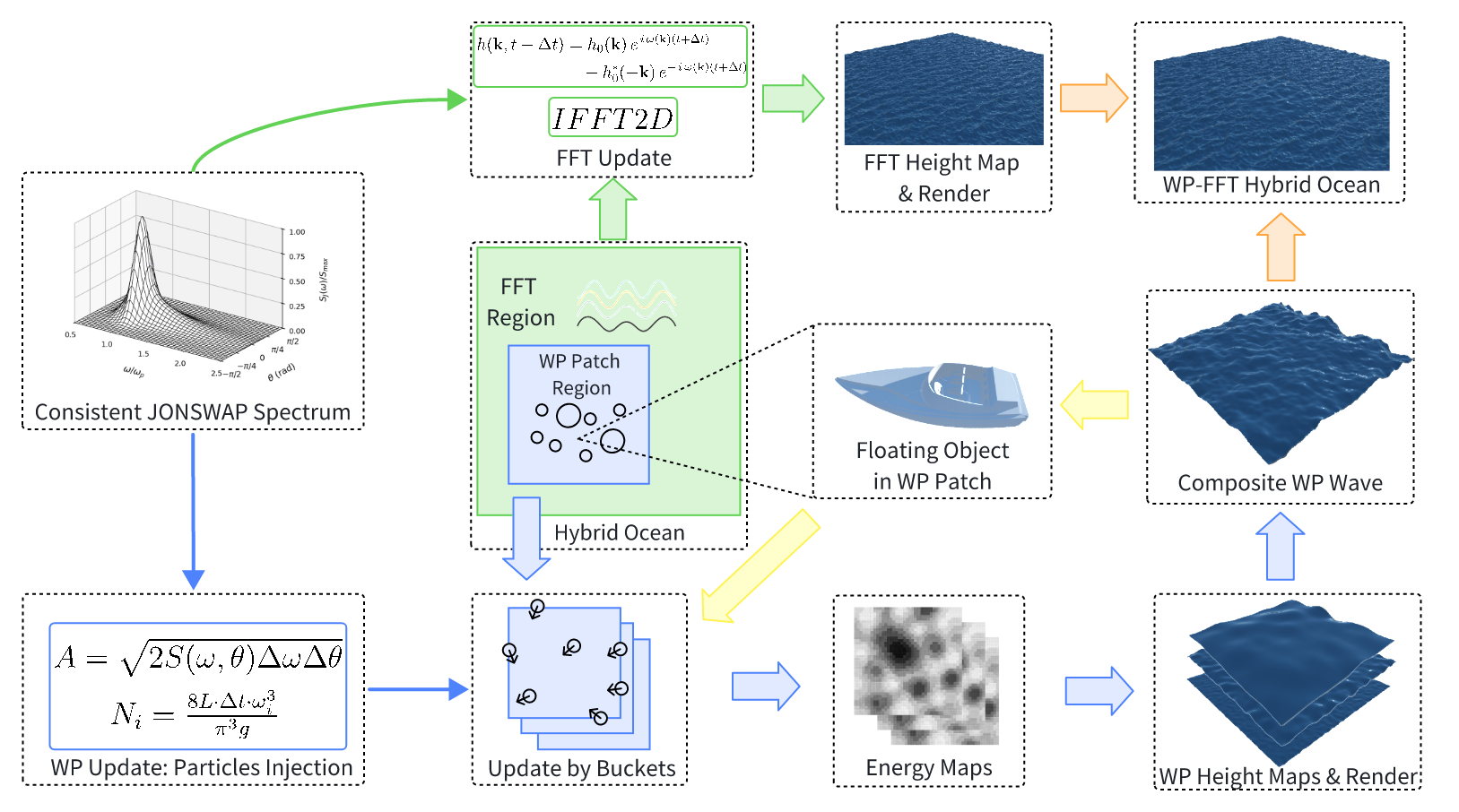} 
  \caption{\textbf{Pipeline overview.} Green arrows indicate FFT updates, 
  blue arrows indicate wave particle (WP) updates, orange arrows 
  indicate coupling between FFT and WP, and yellow arrows indicate 
  interactions between floating objects and WP patches. Both 
  branches are driven by the same spectrum\cite{frechot2006realistic}.}
  \label{fig:pipeline}
\end{figure}
Each frame update follows the pipeline shown in Fig~\ref{fig:pipeline}.
At a high level, both the global FFT field and local WP patches are updated 
under the same spectrum, then combined into the final hybrid ocean.

\textbf{FFT updates} (green arrows) follow the conventional spectral evolution 
and inverse FFT to produce far-field height maps.

\textbf{WP updates} (blue arrows) inject and advect particles inside local patches, 
and their contributions are accumulated into energy and height maps.

\textbf{Coupling} (orange arrows) blends the WP patches with the FFT background 
through smooth interpolation, yielding a seamless hybrid ocean.

\textbf{Object interaction} (yellow arrows) lets floating bodies exchange 
forces with WP patches, generating wakes and ripples while responding to ocean motion.

As FFT updates are standard\cite{tessendorf2001simulating} and interpolation 
is straightforward, we focus on the design of WP updates and 
object interaction, which constitute the novel part of our 
pipeline.

\subsection{Discretization and Rendering}
The frequency domain is discretized into $N_{\omega}$ buckets of width $\Delta\omega$, 
so particles in the same bucket share velocity, radius, and kernel, enabling 
coherent GPU processing. Directions are sampled from the same directional 
spectrum $S(\omega,\theta)$ as the FFT field, amplitudes use Eq.(\ref{eq:A}), 
and the number of particle groups per step $N_i$ follows Eq.(\ref{eq:result}) 
to preserve statistical energy consistency.  

During synthesis, particles accumulate contributions into layered textures 
(one per bucket). Each layer is smoothed by two separable passes along $x$ 
and $y$, then all layers are summed to form the patch height map. Normals 
and displacements are derived by finite differences. Finally, patch maps 
are blended with the FFT background using smooth distance-based weights, 
eliminating seams while preserving global continuity and local detail.

\subsection{Bidirectional Interaction}
To enable two-way coupling, we introduce a spectrum-consistent WP 
emission mechanism driven by object-water interactions.  

On the \emph{object-receives-water} side, buoyancy is computed from 
the blended height and normal maps at probe points, producing drift 
under wind waves.  

On the \emph{object-acts-on-water} side, displacement and suction 
effects are estimated from contact volume and relative velocity, 
and mapped to WP emissions: vertical motions trigger ring ripples, 
while horizontal components yield Kelvin-like wakes. The total 
energy of emitted particles equals the displaced water volume. 
For efficiency, each new particle is assigned to the frequency 
bucket with the closest radius, avoiding the creation of extra 
height-map layers and thus reducing rendering cost.

\section{Experiments and Results}
\subsection{Experimental Setup}
All experiments were conducted on a desktop workstation. The hardware 
and software configuration is as table (\ref{tab:env}).
\begin{table}[!htb]
  \centering
  \caption{Experimental setup}
  \label{tab:env}
  \begin{tabular*}{0.66\linewidth}{@{\extracolsep{\fill}} ll @{}}
    \hline
    CPU    & 13th Gen Intel(R) Core(TM) i7-13700K \\
    GPU    & NVIDIA GeForce RTX 4080 \\
    Memory & 64\,GB DDR5@6400\,MHz \\
    Engine & Unity 2022.3.26f1 \\
    \hline
  \end{tabular*}
\end{table}

\subsection{Results and Performance with Different Parameters}
\begin{figure}[!htb]
  \centering
  \begin{minipage}[t]{1\linewidth}
    \centering
    \includegraphics[width=\linewidth]{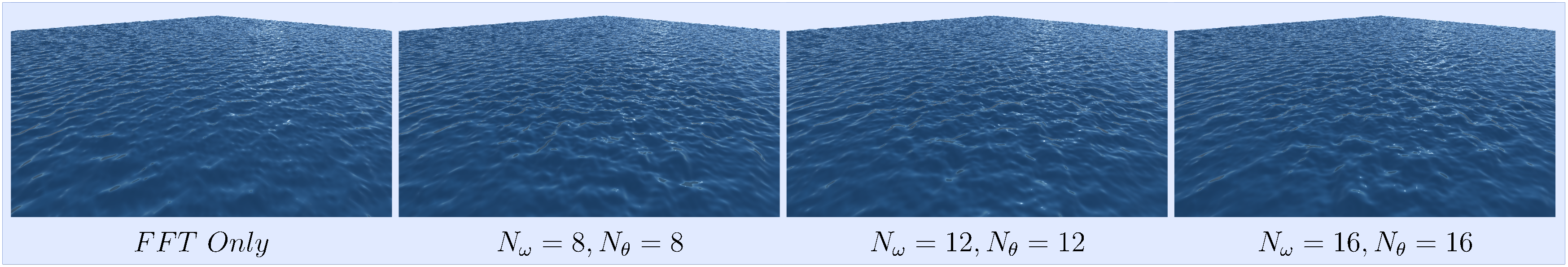}
    \medskip
    \small (a) Results with different sample counts ($N_{\omega},N_{\theta}$)
  \end{minipage}\hfill
  \begin{minipage}[t]{1\linewidth}
    \centering
    \includegraphics[width=\linewidth]{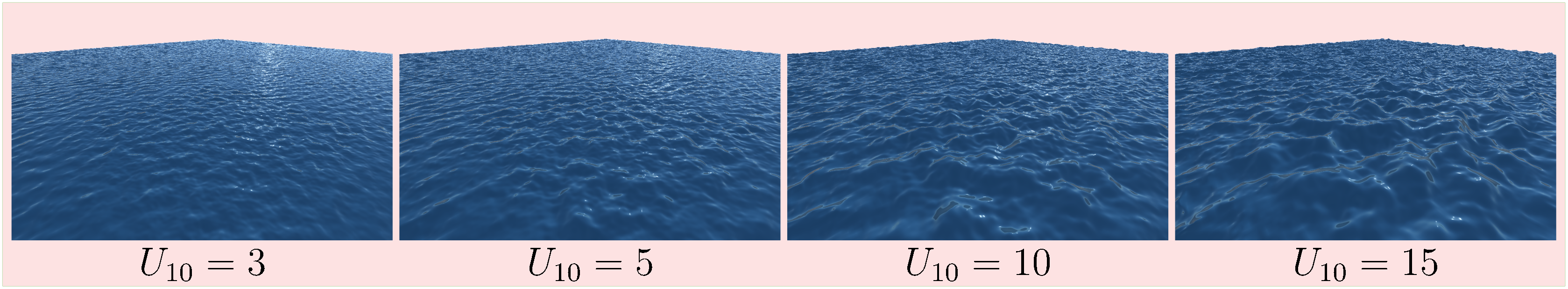}
    \medskip
    \small (b) Results with different wind speeds ($U_{10}$)
  \end{minipage}
    \begin{minipage}[t]{1\linewidth}
    \centering
    \includegraphics[width=\linewidth]{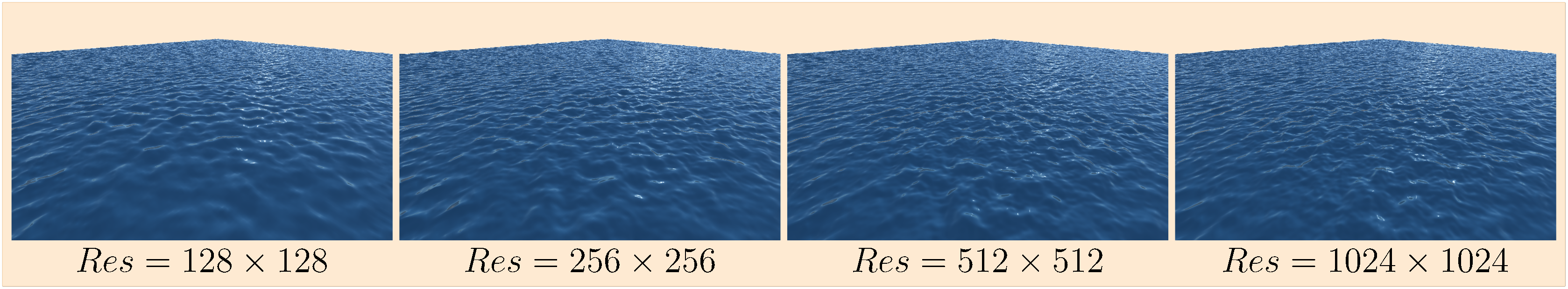}
    \medskip
    \small (c) Results with different WP-Patch resolutions ($Res$)
  \end{minipage}
  \caption{Results with different parameters}
  \label{fig:results_with_parameters}
\end{figure}

To assess the system's robustness under parameter variations, 
we evaluate performance and consistency along three axes: 
sampling resolution ($N_{\omega},N_{\theta}$), wind speed at 
10 meters above the sea ($U_{10}$), and the resolution of the wave particle 
region ($Res$). The hybrid ocean results under different 
parameter settings are shown in Fig.(\ref{fig:results_with_parameters}),
Performance (FPS) is reported in Table.(\ref{tab:performance}).
Unless otherwise specified, the default parameter values are:

\begin{itemize}
  \item $N_{\omega} = 16, \quad N_{\theta} = 16$
  \item $U_{10} = 5, \quad FetchSize = 10000$  
  \item $Res = 512 \times 512$
  \item $OceanSize_{FFT} = 500 \times 500, \quad OceanSize_{WP} = 100 \times 100$
\end{itemize}

\begin{table}[!htb]
  \centering
  \caption{Performance with different parameters}
  \label{tab:performance}
  \begin{tabular*}{0.9\linewidth}{@{\extracolsep{\fill}} |l|l|l|l|l| @{}}
    \hline
    \textbf{Method} & \textbf{$(N_\omega,N_\theta)$} & \textbf{$U_{10}$} & \textbf{$Res$} & \textbf{FPS} \\
    \hline
    FFT-Only     & \textbackslash & 5  & $512 \times 512$ & 2000+ \\
    WP-Only      & (16,16)        & 5  & $512 \times 512$ & 4 \\
    Hybrid (Ours)& (16,16)        & 5  & $512 \times 512$ & 86 \\
    Hybrid (Ours)& (12,12)        & 5  & $512 \times 512$ & 139 \\
    Hybrid (Ours)& (8,8)          & 5  & $512 \times 512$ & 306 \\
    Hybrid (Ours)& (16,16)        & 5  & $256 \times 256$ & 89 \\
    Hybrid (Ours)& (16,16)        & 5  & $1024 \times 1024$ & 56 \\
    Hybrid (Ours)& (16,16)        & 3  & $512 \times 512$ & 39 \\
    Hybrid (Ours)& (16,16)        & 10 & $512 \times 512$ & 220 \\
    Hybrid (Ours)& (16,16)        & 15 & $512 \times 512$ & 318 \\
    \hline
  \end{tabular*}
\end{table}

\paragraph{Sample Count ($N_{\omega},N_{\theta}$)}
Increasing $N_{\omega}$ and $N_{\theta}$ refines frequency-direction 
discretization, producing more particles and richer local detail. 
Since amplitudes follow Eqs.~\ref{eq:A} and \ref{eq:result}, 
energy is redistributed rather than amplified, reducing spectral 
bias. Particle generation and accumulation cost grows roughly 
linearly with sampling, but frequency bucketing keeps it manageable; 
visual improvements saturate beyond $(16,16)$.

\paragraph{Wind Speed ($U_{10}$)}
As $U_{10}$ rises, wave height increases while JONSWAP concentrates 
energy near $\omega_p$, suppressing high-frequency tails. Fewer 
short-wave particles are spawned, lowering per-frame cost; performance 
therefore improves at higher winds without losing near- or far-field fidelity.

\paragraph{Patch Resolution ($Res$)}
Higher $Res$ yields more accurate energy maps but increases filtering cost. 
Below $512^2$, overhead is negligible; above $1024^2$, cost rises sharply 
while visual gains remain minor.

\bigbreak
Overall, different platforms can choose $(N_\omega,N_\theta)$ and $Res$ 
for the desired balance of quality and speed. Across tested settings, 
WP-FFT surfaces remain spectrally consistent, and the propagation 
speed of dominant long waves matches between WP and FFT solutions, 
even if this is difficult to show in still figures.

\subsection{Interactions between Different Objects and Ocean}
\begin{figure}[!htb]
  \centering
  \begin{minipage}[t]{1\linewidth}
    \centering
    \includegraphics[width=\linewidth]{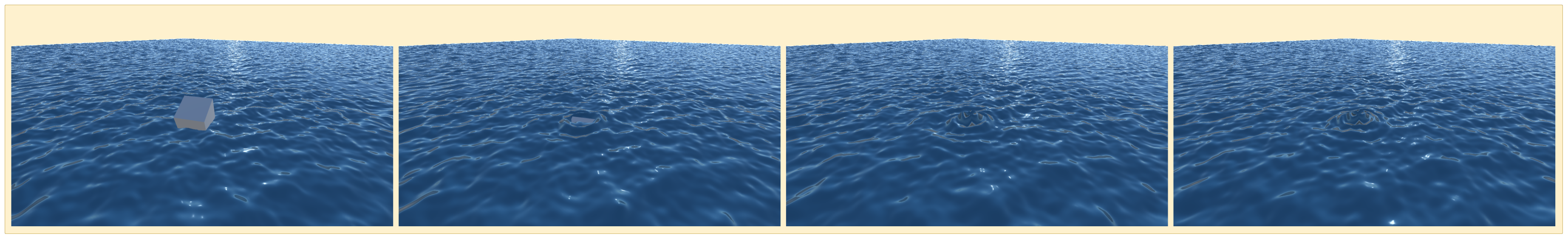}
    \medskip
    \small (a) Interactions between Simple Cube and Ocean
  \end{minipage}\hfill
  \begin{minipage}[t]{1\linewidth}
    \centering
    \includegraphics[width=\linewidth]{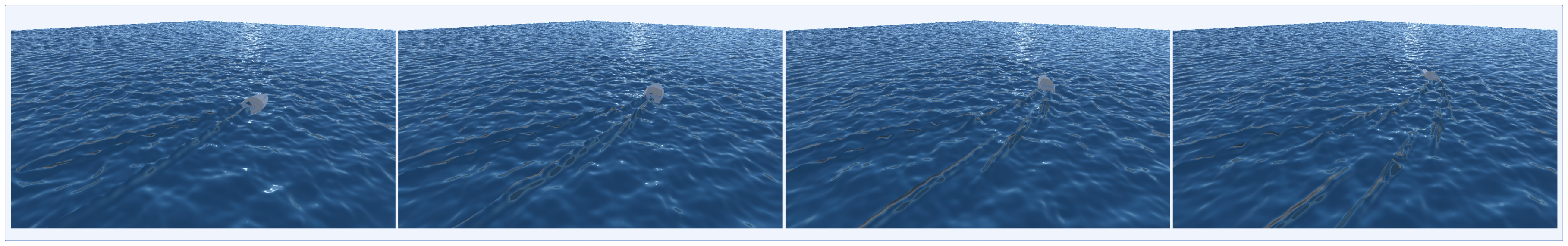}
    \medskip
    \small (b) Interactions between Boat and Ocean
  \end{minipage}\hfill
  \caption{Interactions between Different Objects and Ocean}
  \label{fig:interaction_results}
\end{figure}

As illustrated in Fig.~\ref{fig:interaction_results}, our system 
supports realistic bidirectional interactions between rigid props, 
boats, and the hybrid ocean surface. Objects respond to ocean 
forces through buoyancy drift, while at the same 
time generating ripples and wakes consistent with their volume and 
velocity. 

This demonstrates a key advantage of our hybrid model: 
unlike FFT-only oceans, which cannot represent wakes/ripples, 
or WP-only methods whose large-scale simulation is constrained 
by performance limitations, our method enables localized, 
spectrum-consistent interactions embedded in a large-scale ocean.

\section{Conclusions and Future Work}

\subsection{Conclusions}
We presented a real-time hybrid ocean framework that couples a global FFT background 
with local wave-particle (WP) patches under a unified spectrum. The framework 
introduces two key innovations: (i) a spectrum-consistent WP-FFT representation that 
seamlessly integrates large-scale spectral stability with localized interactions, and 
(ii) an efficient frequency-bucketed sampling and GPU-parallel synthesis scheme that 
maintains spectral consistency while sustaining real-time performance.  

Extensive experiments under varying spectral and sampling parameters validate the 
effectiveness of our approach: the system consistently reproduces coherent wakes and 
ripples while preserving near-far continuity and interactive performance. These 
results demonstrate that the proposed method bridges the gap between large-scale 
spectral realism and fine-grained interactivity in real time.

\subsection{Limitations and Future Work}
Although effective, our framework has two main limitations. 
First, information exchange between the FFT background and WP patches is incomplete: 
ripples generated in WP regions do not propagate back to the far-field FFT surface, 
and new particles are not sampled from the instantaneous FFT boundary state, 
which can reduce cross-boundary coherence. 
Second, the system lacks adaptive sampling: parameters such as $(N_\omega, N_\theta)$ 
and patch resolution remain fixed, rather than adjusting to hardware capability or 
environmental conditions (e.g., wind speed), leading to uneven performance across scenarios.

Future work will focus on addressing these issues, by designing a truly bidirectional 
boundary coupling mechanism that exchanges energy and phase information between FFT and WP, 
and developing adaptive sampling strategies that dynamically balance visual fidelity 
and frame rate across different devices and environments.

\subsubsection{\ackname} 
We thank the members of the DALab at Shanghai 
Jiao Tong University for their insightful discussions 
and continuous support throughout this work. 
\subsubsection{\discintname}
The authors declare no conflicts of interest.

\bibliographystyle{splncs04}
\bibliography{refs}

\begin{thebibliography}{10}
\providecommand{\url}[1]{\texttt{#1}}
\providecommand{\urlprefix}{URL }
\providecommand{\doi}[1]{https://doi.org/#1}

\bibitem{agrotis2016fluid}
Agrotis, A.: „a fluid implicit particle (flip) solver built in houdini,“. Bournemouth University, NCCA  (2016)

\bibitem{bowles2017crest}
Bowles, H., Zimmermann, D., Noris, G., Wang, B.: Crest: Novel ocean rendering techniques in an open source framework. In: Siggraph 2017 Advances in Real-Time Rendering in Games (2017)

\bibitem{donatini2024physically}
Donatini, L., Verwilligen, J., Delefortrie, G., Vantorre, M., Lataire, E.: Physically accurate real-time synthesis of ocean waves for maritime simulators. Applied Ocean Research  \textbf{143},  103866 (2024)

\bibitem{flugge2017realtime}
Fl{\"u}gge, F.J.: Realtime gpgpu fft ocean water simulation  (2017)

\bibitem{forristall1998worldwide}
Forristall, G., Ewans, K.: Worldwide measurements of directional wave spreading. Journal of Atmospheric and Oceanic Technology - J ATMOS OCEAN TECHNOL  \textbf{15} (04 1998). \doi{10.1175/1520-0426(1998)015<0440:WMODWS>2.0.CO;2}

\bibitem{fournier1986simple}
Fournier, A., Reeves, W.T.: A simple model of ocean waves. In: Proceedings of the 13th annual conference on Computer graphics and interactive techniques. pp. 75--84 (1986)

\bibitem{frechot2006realistic}
Fr{\'e}chot, J.: Realistic simulation of ocean surface using wave spectra. In: Proceedings of the first international conference on computer graphics theory and applications (GRAPP 2006). pp. 76--83 (2006)

\bibitem{horvath2015empirical}
Horvath, C.J.: Empirical directional wave spectra for computer graphics. In: Proceedings of the 2015 Symposium on Digital Production. pp. 29--39 (2015)

\bibitem{huang2021ships}
Huang, L., Qu, Z., Tan, X., Zhang, X., Michels, D.L., Jiang, C.: Ships, splashes, and waves on a vast ocean. ACM Transactions on Graphics (TOG)  \textbf{40}(6),  1--15 (2021)

\bibitem{jeschke2018water}
Jeschke, S., Sk{\v{r}}ivan, T., M{\"u}ller-Fischer, M., Chentanez, N., Macklin, M., Wojtan, C.: Water surface wavelets. ACM Transactions on Graphics (TOG)  \textbf{37}(4),  1--13 (2018)

\bibitem{jeschke2017water}
Jeschke, S., Wojtan, C.: Water wave packets. ACM Transactions on Graphics (TOG)  \textbf{36}(4),  1--12 (2017)

\bibitem{kythe2020introduction}
Kythe, P.K.: An introduction to boundary element methods. CRC press (2020)

\bibitem{macklin2013position}
Macklin, M., M{\"u}ller, M.: Position based fluids. ACM Transactions on Graphics (TOG)  \textbf{32}(4),  1--12 (2013)

\bibitem{mastin2007fourier}
Mastin, G.A., Watterberg, P.A., Mareda, J.F.: Fourier synthesis of ocean scenes. IEEE Computer graphics and Applications  \textbf{7}(3),  16--23 (2007)

\bibitem{mazzaretto2022global}
Mazzaretto, O.M., Men{\'e}ndez, M., Lobeto, H.: A global evaluation of the jonswap spectra suitability on coastal areas. Ocean Engineering  \textbf{266},  112756 (2022)

\bibitem{mitsuyasu1975observations}
Mitsuyasu, H., Tasai, F., Suhara, T., Mizuno, S., Ohkusu, M., Honda, T., Rikiishi, K.: Observations of the directional spectrum of ocean wavesusing a cloverleaf buoy. Journal of Physical Oceanography  \textbf{5}(4),  750--760 (1975)

\bibitem{muller2003particle}
M{\"u}ller, M., Charypar, D., Gross, M.: Particle-based fluid simulation for interactive applications. In: Proceedings of the 2003 ACM SIGGRAPH/Eurographics symposium on Computer animation. pp. 154--159 (2003)

\bibitem{newell2008role}
Newell, A.C., Zakharov, V.E.: The role of the generalized phillips' spectrum in wave turbulence. Physics Letters A  \textbf{372}(23),  4230--4233 (2008)

\bibitem{pan2025simulation}
Pan, F., Zou, L., Dong, D., Xiao, S.: Simulation of ocean waves with spectrum-based wave particle. Computer Animation and Virtual Worlds  \textbf{36}(2),  e70014 (2025)

\bibitem{stewart2008introduction}
Stewart, R.H.: Introduction to physical oceanography  (2008)

\bibitem{tessendorf2001simulating}
Tessendorf, J., et~al.: Simulating ocean water. Simulating nature: realistic and interactive techniques. SIGGRAPH  \textbf{1}(2), ~5 (2001)

\bibitem{Tian_Perlin_Choi_2011}
Tian, Z., Perlin, M., Choi, W.: Frequency spectra evolution of two-dimensional focusing wave groups in finite depth water. Journal of Fluid Mechanics  \textbf{688},  169–194 (2011). \doi{10.1017/jfm.2011.371}

\bibitem{yuksel2007wave}
Yuksel, C., House, D.H., Keyser, J.: Wave particles. ACM Transactions on Graphics (TOG)  \textbf{26}(3),  99--es (2007)

\end{thebibliography}

\end{document}